\documentclass[aps,prb,superscriptaddress,twocolumn,showpacs,preprintnumbers,amsmath,amssymb]{revtex4-2}

\usepackage{graphicx,color}
\usepackage{dcolumn}
\usepackage{bm}
\usepackage{ulem}
\usepackage{soul}
\usepackage{hyperref}

\begin{document}

\title{Thickness-induced crossover from strong to weak collective pinning \\
in exfoliated FeTe$_{0.6}$Se$_{0.4}$ thin films at 1~T}

\author{Ryoya~Nakamura}
\affiliation{Department of Physics, Graduate School of Science, Osaka University, Toyonaka 560-0043, Japan}
\author{Masashi~Tokuda}
\affiliation{Department of Physics, Graduate School of Science, Osaka University, Toyonaka 560-0043, Japan}
\author{Mori~Watanabe}
\affiliation{Department of Physics, Graduate School of Science, Osaka University, Toyonaka 560-0043, Japan}
\author{Masamichi~Nakajima}
\affiliation{Department of Physics, Graduate School of Science, Osaka University, Toyonaka 560-0043, Japan}
\author{Kensuke~Kobayashi}
\affiliation{Department of Physics, Graduate School of Science, Osaka University, Toyonaka 560-0043, Japan}
\affiliation{Institute for Physics of Intelligence and Department of Physics, Graduate School of Science, The University of Tokyo, Tokyo 113-0033, Japan}
\author{Yasuhiro~Niimi}
\email{niimi@phys.sci.osaka-u.ac.jp}
\affiliation{Department of Physics, Graduate School of Science, Osaka University, Toyonaka 560-0043, Japan}
\affiliation{Center for Spintronics Research Network, Osaka University, Toyonaka 560-8531, Japan}

\begin{abstract}
We studied flux pinning in exfoliated FeTe$_{0.6}$Se$_{0.4}$ thin-film devices with 
a thickness $d$ from 30 to 150~nm 
by measuring the critical current density $J_{\mathrm{c}}$. 
In bulk FeTe$_{0.6}$Se$_{0.4}$, the flux pinning has been discussed in the framework of 
weak collective pinning, while there is little knowledge on the pinning mechanism in the
thin-film region. From the thickness $d$ dependence of $J_{\mathrm{c}}$ 
at a fixed magnetic field of 1~T, 
we found that the strong pinning is dominant below $d \approx 70$~nm, while 
the weak collective pinning becomes more important above $d \approx 100$~nm. 
This crossover thickness can be explained by the theoretical model 
proposed by van der Beek \textit{et al} [Phys. Rev. B. {\bf 66}, 024523 (2002)]. 
\end{abstract}

\maketitle

\section{\label{sec:level1}Introduction}
Since 2008, iron-based superconductors (IBSs) have been well studied 
as a new group of high-temperature superconductors 
because of the two-dimensional layered structure similar to 
cuprate superconductors~\cite{Kamihara2008}. 
Their fundamental properties such as
the short coherence length and large upper-critical field 
are useful for 
superconducting wires~\cite{Hosono2018}. 
For such applications,  
it is essential to increase the critical current density $J_{\mathrm{c}}$.
In order to achieve this,
the pinning mechanism of magnetic vortices needs to be clarified.

Among all kinds of IBSs, the iron-based chalcogenide FeTe$_{1-x}$Se$_x$ with $x = 0.4\sim0.5$ (FTS) 
has the simplest crystal structure with only 
Fe$\mathit{Ch}$ ($\mathit{Ch} = $~Se or Te) layers. 
There have been many reports on 
bulk~\cite{Sales2009, Das2011, Bonura2012, Galluzzi2020, Sun2013, Tamegai2016, Yadav2011, Miu2012, Shahbazi2013JoAP, Wu2016} and thin-film~\cite{Iida2013, Braccini2013, Yuan2016, Bryja2017} 
FTS single crystals, 
which enable a detailed discussion of the pinning mechanism. 
Although the critical temperature $T_{\mathrm{c}}$ is not so high, 
the upper critical field is comparable to other IBSs.  
Thus, FTS has a potential to replace low-temperature superconducting wires 
such as Nb-Ti superconductors. 
In addition, FTS is a candidate for topological superconductor. 
Particularly it has been intensively studied on the detection of 
zero energy vortex bound state (ZVBS) in magnetic vortices, 
which is a fingerprint of the Majorana quasiparticles~\cite{Zhang2018,Wang2018,Machida2019,Zhu2020}. 
In fact, a recent scanning tunneling microscopy study demonstrated 
the existence of ZVBS in magnetic vortices~\cite{Machida2019}. 
Simultaneously, however, it also revealed that some of the magnetic vortices 
do not contain the ZVBS. This suggests that 
the Majorana zero modes gain finite energies via the interaction between 
magnetic vortices, resulting in an energy splitting of the zero modes. 
In order to unveil the ZVBS in the superconducing FTS, 
it is helpful to understand the magnetic vortex lattice and also 
its pinning mechanism under finite magnetic fields.  

In general, there are two pinning mechanisms in superconductors, 
i.e., weak collective pinning~\cite{Blatter1994, Griessen1994} 
and strong pinning~\cite{Ovchinnikov1991, Beek2002}. 
In the former case, the pinning stems from 
the atomic-scale inhomogeneity of the superconducting regions 
due to impurities or defects 
and is further categorized into two types. 
One is $\delta l$ pinning, originating from spatial fluctuations of the mean free path $l$ 
due to lattice defects~\cite{Blatter1994}. 
This is related to the derivative of the macroscopic wave function 
in the Ginzburg-Landau theory.
The other is $\delta T_{\mathrm{c}}$ pinning, 
originating from spatial fluctuations of the critical 
temperature $T_{\mathrm{c}}$~\cite{Blatter1994}. 
This is related to the fact that the coefficient $\alpha$ 
of the probability density for the macroscopic wave function in the Ginzburg-Landau theory 
is proportional to $T_{\mathrm{c}}-T$. 
For FTS single crystals, 
the pinning mechanisms have been 
determined by analyzing magnetization measurements 
with the Bean model~\cite{Das2011, Bonura2012, Galluzzi2020, Sun2013, Tamegai2016}. 
In most cases the $\delta l$ pinning is 
dominant~\cite{Das2011, Bonura2012, Galluzzi2020}, 
while in some cases both the $\delta l$ and $\delta T_{\mathrm{c}}$ pinnings 
coexist~\cite{Sun2013, Tamegai2016}. 

Contrary to the weak collective pinning, 
the strong pinning takes place at large-sized defects 
comparable to the coherence length ($\approx 10$~nm)
and can induce a high critical current density~\cite{Yeoh2004,Chen2009}. 
It has been established that the strong pinning is a dominant mechanism 
for 100--300~nm thick YBa$_{2}$Cu$_{3}$O$_{7-x}$ (YBCO) films where 
the critical current density has a large value~\cite{Beek2002,Ijaduola2006}. 
Although there are some reports on 
enhanced critical current densities~\cite{Mele2012,Si2013,Lee2017} 
as well as strong pinning~\cite{Iida2010,Zhang2011,Bellingeri2012} in IBS thin films, 
the relation between the strong pinning and the thickness of IBSs has not been fully 
elucidated yet. 

In this work, 
we directly obtained $J_{\mathrm{c}}$ in exfoliated FeTe$ _{0.6} $Se$ _{0.4}$ 
thin-film devices 
with several different thicknesses $d$
by measuring current-voltage properties
at a fixed out-of-plane magnetic field of 1~T. 
For $d \lesssim 70$~nm, $J_{\mathrm{c}}$ 
linearly increases with increasing $d$. 
Such a tendency is consistent with the strong pinning reported in YBCO 
thin-film superconductors. 
When $d$ exceeds $\approx 100$~nm, 
$J_{\mathrm{c}}$ 
decreases with increasing $d$ and eventually approaches 
the value estimated for bulk FTS. 
By plotting $J_{\mathrm{c}}$ as a function of $d$, 
we observe a crossover behavior from the strong pinning 
in thin-film regions to the weak collective pinning. 
This crossover thickness $d^* = 70 \sim 100$~nm 
can be explained by the theoretical model 
proposed by van der Beek \textit{et al}~\cite{Beek2002}.

\section{Experimental details}

\begin{figure}
\begin{center}
\includegraphics[width=6.5cm]{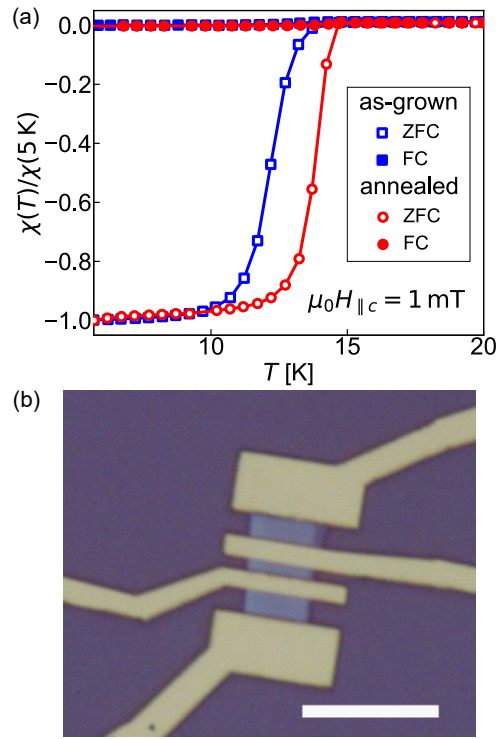}
\caption{ 
(a) Temperature dependence of magnetic susceptibilities of 
as-grown bulk FTS (square) and annealed bulk FTS (circle) 
under the magnetic field ($\mu_{0}H_{\parallel c}$)
of 1~mT. 
The closed and open symbols 
indicate data obtained with field cooling (FC) and ZFC, respectively. 
The vertical axis is normalized by the data for ZFC at $T=5$~K for each sample. 
(b) Optical microscope image of our typical device.
The white scale bar corresponds to 10 $\mathrm{\mu m}$.}
\label{fig:1}
\end{center}
\end{figure}

Single crystals of FeTe$ _{0.6} $Se$ _{0.4}$ were grown from a stoichiometric mixture 
of Fe, Te, and Se powder. We note that in the present work, 
FeTe$ _{0.6} $Se$ _{0.4}$ has been studied, but we use the same abbreviation (FTS) as for 
a general composition ratio FeTe$_{1-x}$Se$_x$ in this paper. 
The mixture was loaded into an alumina crucible and sealed 
in an evacuated quartz tube. The quartz tube was heated at 650$^\circ$C for 10 h and then 
1070$^\circ$C for 15 h, followed by cooling down to 620$^\circ$C at a rate of 3$^\circ$C/h. 
FTS has a van-der-Waals interaction between 
the two adjacent \textit{Ch} layers, which makes it easier to fabricate thin-film devices using 
the mechanical exfoliation technique. On the other hand, it has been known that 
excess Fe is present in between the \textit{Ch} layers,  
which gives rise to magnetic correlations and suppresses the superconductivity~\cite{Bendele2010}. 
To remove the excess Fe, we annealed as-grown FTS bulk crystals at 400$^\circ$C 
for 40 h under 1\% atmosphere of O$_2$ gas~\cite{sun_2013,Sun2014,Sun2019}. 
To determine $T_{\mathrm{c}}$ of FTS bulk crystals, 
we measured the $dc$ magnetic susceptibility using 
Magnetic Property Measurement System (Quantum Design). 
Figure~\ref{fig:1}(a) shows the temperature dependence of the magnetic susceptibility $\chi$ 
for as-grown and annealed FTS. 
In this work, $T_{\mathrm{c}}^{\mathrm{bulk}}$ is defined as a temperature 
where $\chi$ for zero-field cooling (ZFC) starts to decrease. 
From the $\chi$ measurements, we determine $T_{\mathrm{c}}^{\mathrm{bulk}} = 13.7$~K 
for the as-grown sample and $T_{\mathrm{c}}^{\mathrm{bulk}} = 14.4$~K for the
annealed crystal. By annealing the FTS crystal, $T_{\mathrm{c}}^{\mathrm{bulk}}$ has been 
enhanced by 5\%, which is consistent with Refs.~\cite{sun_2013,Sun2014,Sun2019}.
Hereafter, we mainly focus on the annealed FTS unless otherwise noted, 
whereas we also use the as-grown FTS for comparison.

To obtain thin-film devices, we adopted the mechanical exfoliation technique using Scotch tapes 
under ambient conditions.
Some of the exfoliated FTS flakes onto the Scotch tape 
were transferred to a thermally oxidized silicon substrate. 
We then coated polymethyl-methacrylate resist on the substrate
and patterned electrodes with electron beam lithography.
After the development of the resist,
Ti (5~nm) and Au (150~nm) were deposited. 
Ti works as an adhesion layer for the Si/SiO$_{2}$ substrate. 
Before the deposition of electrodes, 
we performed Ar milling process for 55 s to remove
the residual resist at the surface of FTS. 
In fact, we have confirmed the following: 
(1) This process does not give significant damage to FTS; 
(2) without the Ar milling process, we cannot obtain an Ohmic contact 
between FTS and the electrodes because of the residual resist at the interface.
Figure~\ref{fig:1}(b) shows 
an optical microscope image of a typical device.
The thickness $d$ of FTS was determined with a commercially available 
atomic force microscopy
after finishing transport measurements. 
Although we sometimes found not-perfectly homogeneous thin-film flakes, 
we regarded $d$ as a film thickness when surface regions with $d$ exceeds 85\%. 
In the present work, $d$ ranges from 30 to 150~nm. 
Because of the limitation of the present exfoliation method, 
we could not obtain FTS devices thicker than 150~nm.
We measured the resistivity $\rho$ of thin-film FTS devices with a standard $ac$ Lock-in technique. 
To determine $J_{\mathrm{c}}$, we performed $dc$ current ($I$)-voltage ($V$) measurements.

\begin{figure}
\begin{center}
\includegraphics[width=8.5cm]{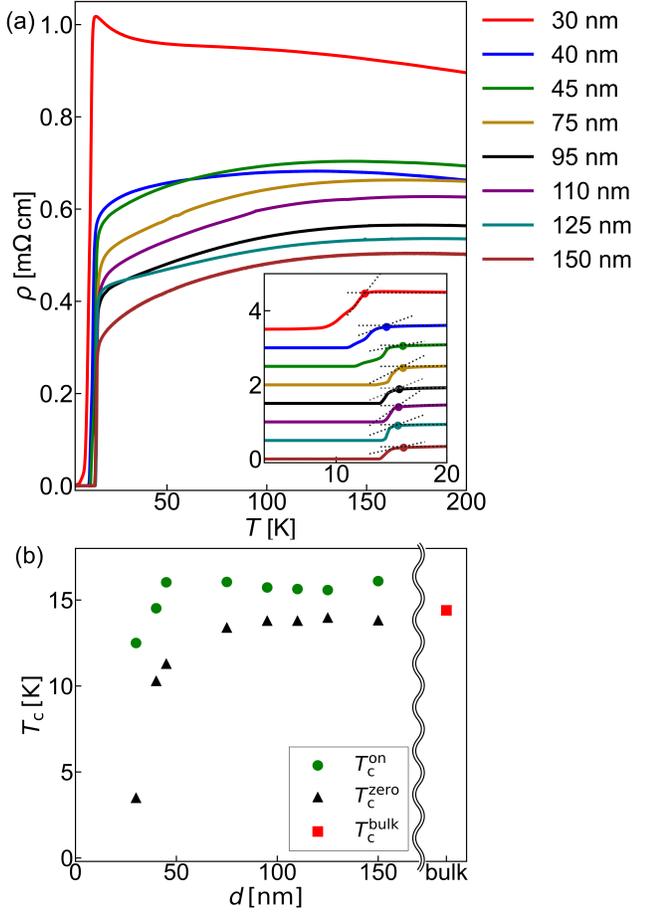}
\caption{ 
(a) Temperature $T$ dependence 
of $\rho$ of FTS thin-film devices.
The inset shows the closeup near $T_{\mathrm{c}}$.
We define two $T_{\mathrm{c}}$, i.e., 
$T_{\mathrm{c}}^{\mathrm{zero}}$ and $T_{\mathrm{c}}^{\mathrm{on}}$, as detailed in the main text.
(b) Thickness $d$ dependence 
of $T_{\mathrm{c}}^{\mathrm{zero}}$ and $T_{\mathrm{c}}^{\mathrm{on}}$. 
We also plot $T_{\mathrm{c}}^{\mathrm{bulk}}$ 
of annealed bulk FTS crystal for reference.}
\label{fig:2}
\end{center}
\end{figure}

\section{experimental results}

\begin{figure}
\begin{center}
\includegraphics[width=8.5cm]{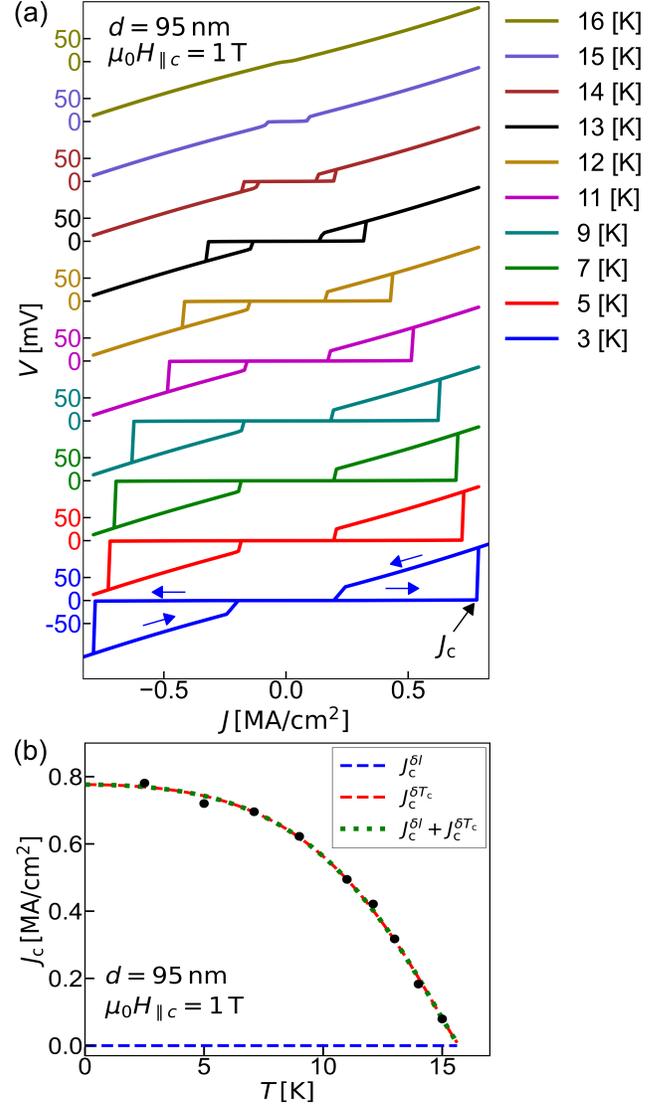}
\caption{ 
(a) Current-voltage properties for $d=95$~nm thick device
under the out-of-plane magnetic field $\mu_0 H_{\parallel c} = 1\,\mathrm{T}$ 
at various temperatures. 
(b) Temperature dependence of $J_{\mathrm{c}}(T)$ for 
 $d=95$~nm thick device under the out-of-plane magnetic field 
$\mu_0H_{\parallel c}=1\, \mathrm{T}$. 
The green dotted line is the best fit with Eqs.~(\ref{eq:2})--(\ref{eq:4}), 
while the red and blue dashed lines are the contributions from 
$\delta T_{\mathrm{c}}$ and $\delta l$ pinnings, respectively.
}
\label{fig:3}
\end{center}
\end{figure}

First, to determine $T_{\mathrm{c}}$ of FTS thin-film devices, we 
measured the temperature $T$ dependence of the resistivity $\rho$ in Fig.~\ref{fig:2}(a). 
When the film thickness $d$ is larger than 100~nm, 
$\rho$ decreases with decreasing $T$ for the whole temperature region.
Below $d \approx 100$~nm, on the other hand, $\rho$ slightly increases with decreasing $T$, 
takes a maximum and becomes zero at low temperatures. 
In particular, a 30~nm thick device shows insulating behavior down to 13~K.
Such tendencies are due to the inhomogeneity of 
superconducting states in FTS crystals, as reported in Ref.~\cite{Yue2016}. 
The superconducting percolation network, 
which is strongly connected in bulk, 
gradually weakens with decreasing $d$. 
This results in weak superconductivity for thinner FTS film devices. 
Figure~\ref{fig:2}(b) shows the thickness dependence of $T_{\mathrm{c}}$. 
There are two ways to define $T_{\mathrm{c}}$: 
One is to use the initial rise of resistivity from zero ($T_{\mathrm{c}}^{\mathrm{zero}}$),  
and the other is to use the onset of resistivity drop in $\rho$-$T$ curve 
($T_{\mathrm{c}}^{\mathrm{on}}$) as indicated in Fig.~\ref{fig:2}(a). 
Since there is a finite voltage jump 
even just below $T_{\mathrm{c}}^{\mathrm{on}}$ [see Fig.~\ref{fig:3}(a)], 
in this work we have adopted $T_{\mathrm{c}}^{\mathrm{on}}$ as $T_{\mathrm{c}}$, 
namely, $T_{\mathrm{c}} \equiv T_{\mathrm{c}}^{\mathrm{on}}$. 
$T_{\mathrm{c}}$ starts to decrease below $d \lesssim 100$~nm and 
should vanish at $10 \sim 20$~nm. According to Ref.~\cite{Sun2014}, 
the critical thickness $d_{0}$ below which $T_{\mathrm{c}}$ vanishes is 12~nm, 
which is consistent with the present work. 

Next we performed $I$-$V$ measurements 
under the out-of-plane magnetic field 
$\mu_0H_{\parallel c}=1$~T.
Figure~\ref{fig:3}(a) shows $dc$ current-voltage curves 
for a 95~nm thick FTS device at several different temperatures. 
We use the current density $J$ instead of $I$ to compare with different $d$ devices. 
At low temperatures, there is a clear hysteresis that originates from the Joule heating 
in the FTS device. In the present work, 
we define $J_{\mathrm{c}}$ as the current density 
when the measured $dc$ voltage exceeds a threshold value of 1~mV.
For example, $J_{\mathrm{c}}$ is 0.78~MA/cm$^{2}$ at $T = 3$~K. 
In the vicinity of $T_{\mathrm{c}}$, on the other hand, the hysteresis vanishes and 
the voltage jump near $J_{\mathrm{c}}$ becomes less clear.  
We also performed pulse current measurements with a pulse width of 1~ms and an interval of 100~ms 
in order to evaluate $J_{\mathrm{c}}$ without the effects of Joule heating.
The obtained $J_{\mathrm{c}}$ was almost the same value 
as in Fig.~\ref{fig:3}(a).
In Fig.~\ref{fig:3}(b), $J_{\mathrm{c}}$ obtained from Fig.~\ref{fig:3}(a) is plotted as a function of $T$. 
$J_{\mathrm{c}}$ monotonically decreases with increasing $T$ 
and vanishes at $T_{\mathrm{c}}$.

\begin{figure}
\begin{center}
\includegraphics[width=7cm]{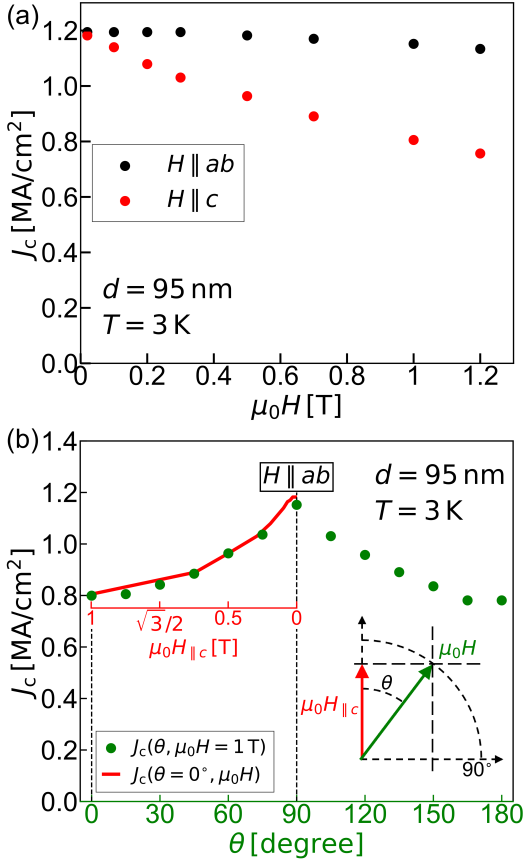}
\caption{ 
(a) $J_{\mathrm{c}}$ of $d=95$~nm thick device measured at $T=3$~K 
as a function of $\mu_{0}H$ along the $ab$ plane (black circle) and 
the $c$ axis (red circle).
(b) $J_{\mathrm{c}}$ of $d=95$~nm thick device measured 
at $T=3$~K and $\mu_{0}H=1$~T
as a function of angle $\theta$ from the $c$ axis (green circle). 
We show the definitions of $\theta$ and $\mu_{0}H$ in the inset. 
The magnetic field applied along the $ab$ plane corresponds to 
$\theta = 90^{\circ}$.
}
\label{fig:4}
\end{center}
\end{figure}

We also measured the magnetic field and angle dependences of 
$J_{\mathrm{c}}$ for the $d=95$~nm thick FTS device at $T=3$~K. 
Figure~\ref{fig:4}(a) shows $J_{\mathrm{c}}$ as a function of $\mu_{0}H$ 
along the $c$ axis and the $ab$ plane. When the magnetic field is applied along 
the $ab$ plane, $J_{\mathrm{c}}$ is more or less constant. 
For the perpendicular magnetic field $(\mu_{0}H\parallel c)$, on the other hand,  
$J_{\mathrm{c}}$ decreases with increasing $\mu_{0}H$. 
Figure~\ref{fig:4}(b) shows $J_{\mathrm{c}}$ at $\mu_{0}H = 1$~T 
as a function of the rotation angle $\theta$ from the $c$ axis. 
$J_{\mathrm{c}}\left( \theta \right) $ has a broad maximum 
at $\theta = 90^{\circ}\,(\mu_{0}H\parallel ab)$,
and there is no peak at $\theta = 0^{\circ}$ nor
$180^{\circ}\,(\mu_{0}H\parallel c)$. 
If we assume that $J_{\mathrm{c}}(\theta,\mu_0H)$ 
depends only on the 
$c$ axis component of the applied magnetic field, i.e., 
$\mu_{0}H_{\parallel c} = \mu_{0}H\cos\theta$, 
the angle dependence of $J_{\mathrm{c}}$ is expected as shown by the red curve, 
which is consistent with the experimental data.  
This clearly shows that $J_{\mathrm{c}}$ depends only on the perpendicular 
component of the magnetic field.
Such magnetic field and angle dependences of $J_{\mathrm{c}}$ are consistent with 
previous works on FTS thin films~\cite{Iida2013, Braccini2013, Yuan2016, Bryja2017}.

\begin{figure}
\begin{center}
\includegraphics[width=7cm]{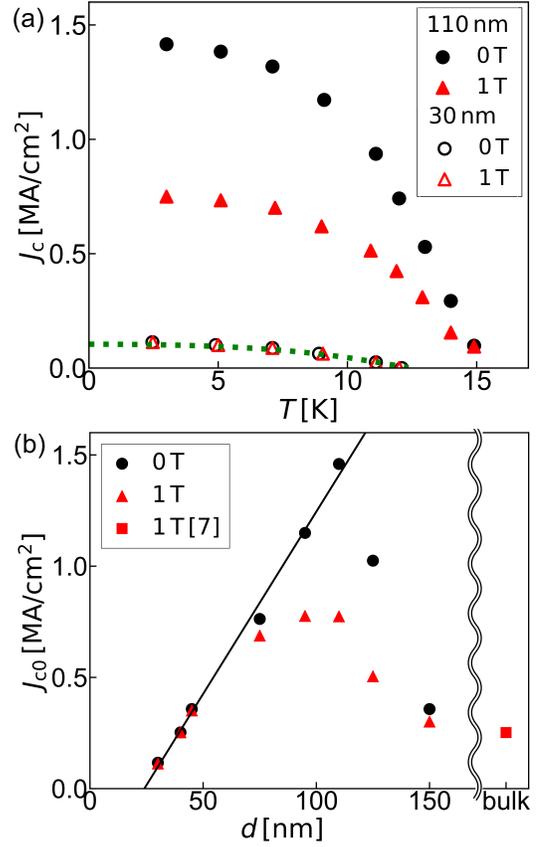}
\caption{ (a) 
Temperature dependence of $J_{\mathrm{c}}(T)$ 
at $\mu_0H_{\parallel c}=0$~T (black circles) and 1~T (red triangles)
for $d=30$ and 110~nm. 
The green dotted line is the best fit for $d=30$~nm at $\mu_{0}H_{\parallel c}=0$~T
using Eqs.~(\ref{eq:2})--(\ref{eq:4}). 
(b) Thickness dependence of 
$J_{\mathrm{c0}}$ for $\mu_{0}H_{\parallel c}=0$~T (black circles)
and $1$~T (red triangles). 
For comparison, we also plot $J_{\mathrm{c}}$ at $\mu_{0}H_{\parallel c}=1$~T 
for bulk FTS in Ref.~\cite{Sun2013} (red square). 
The solid line is the best fit with Eq.~(\ref{eq:1}) 
for $J_{\mathrm{c0}}$ below $d=50$~nm. 
}
\label{fig:5}
\end{center}
\end{figure}

Now we move on to the $d$ dependence of $J_{\mathrm{c}}$. 
In Fig.~\ref{fig:5}(a), we show 
the temperature dependence of $ J_{\mathrm{c}}(T) $ measured at 
$\mu_0H_{\parallel c}=0$~T and 1~T for $d=30$ and 110~nm thick 
devices, respectively.
There is almost no difference in $ J_{\mathrm{c}}(T) $ 
between 0 and 1~T for the 30~nm thick device, 
whereas the difference is significant for the 110~nm thick device. 
To see the $d$ dependence of $ J_{\mathrm{c}}(T) $ at 
0 and 1~T more systematically, 
we plot $J_{\mathrm{c0}}$, which is obtained by extrapolating 
the $J_{\mathrm{c}}(T)$ vs $T$ curves down to $T=0$ as detailed in 
the next section at $\mu_0H_{\parallel c}=0$ and 1~T. 
$J_{\mathrm{c0}}$ linearly increases with $d$ up to $d \approx 70$~nm for 1~T 
and $d \approx 100$~nm for 0~T. 
Above $d \approx 100$~nm, $J_{\mathrm{c0}}$ suddenly decreases with increasing $d$ 
and approaches a $J_{\mathrm{c}}$ value evaluated with the Bean model~\cite{Sun2013}. 
Figure~\ref{fig:5}(b) clearly indicates that there are two regions below $d \approx 70$~nm 
and above $d \approx 100$~nm; the former is the strong pinning region and the latter is 
the weak collective pinning region, as discussed in more detail in the next section.

\section{discussions}

\subsection{Strong pinning}

We first discuss the strong pinning region below $d \approx 70$~nm. 
It is known that 
FTS is a highly inhomogeneous material~\cite{Joseph2010,Hu2011,Singh2013,Zhu2013,Machida2019}. 
According to electron energy loss spectroscopy measurements~\cite{Zhu2013}, 
there is an inhomogeneous distribution of Te in FTS crystals; 
it forms some clusters in FTS and the spatial variation of the cluster is approximately 10~nm.
Therefore, the effect of the inhomogeneity 
should be prominent 
when $d$ is of the order of 10~nm, where the strong pinning is 
more dominant~\cite{Beek2002,Ijaduola2006}. 

Lower magnetic fields 
are favorable to confirm whether the strong pinning is essential in our thin-film FTS devices.
This is because the vortex density in the low field region 
is too small for weak collective pinning to contribute 
and only the strong pinning contributes to $J_{\mathrm{c}}$~\cite{Beek2010}. 
Therefore, we measured $J_{\mathrm{c}}$ at $\mu_{0}H=0$~T 
(but probably with a small Oersted field). 
Then $J_{\mathrm{c0}}$, $J_{\mathrm{c}}$ at $T=0$, was obtained by fitting 
$J_{\mathrm{c}}$ vs $T$ curves with some theoretical model. 
We first tried theoretical expressions discussed in Ref.~\cite{Beek2002} for the strong pinning, 
but none of them could reproduce our experimental data. 
Thus we adopted Eqs.~(\ref{eq:2})--(\ref{eq:4}) to fit the $J_{\mathrm{c}}$ vs $T$ curves. 
We note that Eqs.~(\ref{eq:2})--(\ref{eq:4}) are originally developed 
for the weak collective pinning region, as detailed in the next subsection. 
As shown in Fig.~\ref{fig:5}(a), 
$J_{\mathrm{c}}$ for the $d=30$~nm thick device 
can be fitted with Eqs.~(\ref{eq:2})--(\ref{eq:4}). 
The evaluated $J_{\mathrm{c0}}$ does not depend on the applied magnetic field 
at least within 1~T and linearly increases with increasing $d$. 
In fact, these features cannot be explained by 
the weak collective theory but are consistent with the strong pinning theory, as detailed below. 
This fact also suggests that Eqs.~(\ref{eq:2})--(\ref{eq:4}) would be useful even for the 
strong pinning.

As mentioned above, 
$ J_{\mathrm{c0}} $ at 0~T has the same value as at 1~T and 
linearly increases with $d$ up to $d \approx 100$~nm [see Fig.~\ref{fig:5}(b)].
Such a $d$ dependence of $ J_{\mathrm{c}}(d) $ has also been observed in 
YBCO films~\cite{Sheriff1997,Beek2002,Ijaduola2006,Wang2007,Cherpak2008}.
According to a theoretical model proposed by van der Beek \textit{et al}.~\cite{Beek2002}, 
the critical current density for very thin superconducting films is given by 
\begin{align}
    J_{\mathrm{c0}}=\frac{\varepsilon_0}{4\pi\Phi_0\xi_{ab}}\ln{\left(1+\frac{D_{ab}^2}{2\xi_{ab}^2}\right)}\frac{D_{c}}{{(d^{*}-d_0)}^2}(d-d_0), \label{eq:1}
\end{align}
where $\varepsilon_0=(\Phi_0/4\pi\lambda_{ab})^2(4\pi/\mu_0)$ is 
the typical energy scale for single vortex,
$\Phi_0=h/2e$ is the flux quantum,
$\lambda_{ab}$ is the in-plane magnetic penetration depth at $T=0$, 
$\mu_0=4\pi\times10^{-7}$~H/m, 
$\xi_{ab}$ is the in-plane coherence length at $T=0$, 
and $d_0$ is the critical thickness where $ J_{\mathrm{c0}} $ becomes zero.
$D_{ab}$ and $D_{c}$ are the typical sizes of defect along the $ab$ plane and $c$ axis, 
respectively;
$d^*$ is the crossover film thickness from 
the very thin-film region to relatively thick film region~\cite{Beek2002}. 
As described by Eq.~(\ref{eq:1}), $ J_{\mathrm{c0}} $ 
in the very thin-film region should
be field-independent and linearly increases with increasing $d$. 
This is consistent with the $d$ dependence of $ J_{\mathrm{c0}} $ 
for $d \lesssim 70$~nm in 
Fig.~\ref{fig:5}(b), at least in the field range from 0 to 1~T.
Furthermore, $d_0$ is also consistent with the thickness below 
which $ T_{\mathrm{c}} $ vanishes in Fig.~\ref{fig:2}(b).
The contribution of the strong pinning is more significant 
in this thickness range.

As we increase $d$ further from 70~nm, 
a significant difference in $ J_{\mathrm{c0}} $ is observed between $\mu_{0}H=0$ and 1~T.
$ J_{\mathrm{c0}} $ takes a maximum value at $d \approx 100$~nm 
for both magnetic fields and suddenly decreases with increasing $d$. 
This thickness range will be discussed 
in the next subsection. 
From the linear fitting for $J_{\mathrm{c0}}$ below $d \approx 70$~nm, we obtain the slope 
$J_{\mathrm{c0}}/(d-d_0) \sim 1.6\times 10^{17}~\mathrm{A/m^3}$. 
By using $\xi_{ab}=2.5$~nm~\cite{Iida2013}, 
$\lambda_{ab}=0.49$~$\mu$m~\cite{Bendele2010} and 
assuming $D_{ab} = D_c = 10 \sim 20$~nm, 
we estimate the crossover film thickness $d^{*}$ to be $70 \sim 110$~nm.
This $d^{*}$ value is consistent with the experimental crossover thickness ($70 \sim 100$~nm)
shown in Fig.~\ref{fig:5}(b).

\begin{figure}
\begin{center}
\includegraphics[width=8.5cm]{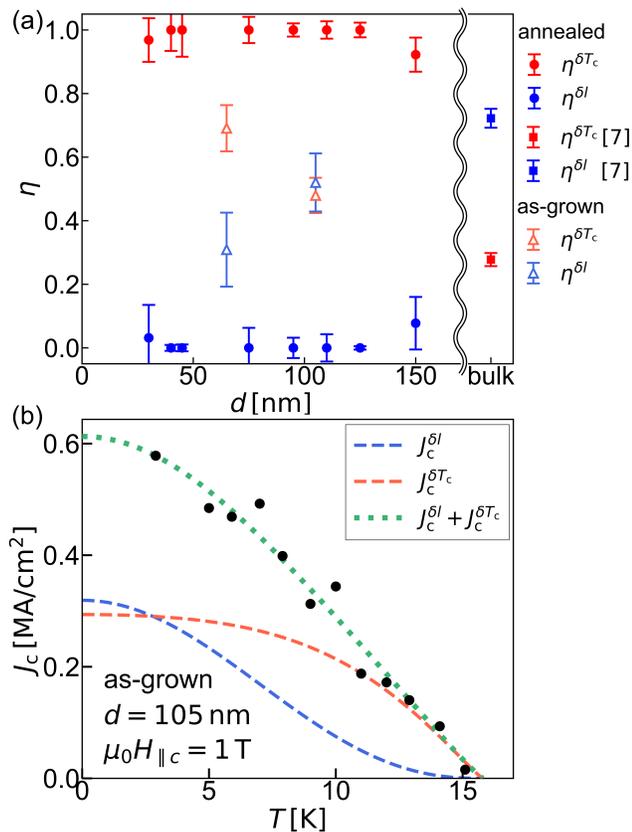}
\caption{ 
(a) Thickness dependence of 
$\eta^{\delta l}$ (blue circles) 
and $\eta^{\delta T_{\mathrm{c}}}$ (red circles) 
for annealed FTS samples. We also plot $\eta^{\delta l}$ (blue triangles) 
and $\eta^{\delta T_{\mathrm{c}}}$ (red triangles) for as-grown FTS samples.
For comparison, we also plot $\eta$ for bulk FTS 
in Ref.~\cite{Sun2013} (square). 
(b) Temperature dependence of $J_{\mathrm{c}}(T)$ for 
$d=105$~nm thick as-grown device under the out-of-plane magnetic field 
$\mu_0H_{\parallel c}=1\, \mathrm{T}$. 
The green dotted line is the best fit with Eqs.~(\ref{eq:2})--(\ref{eq:4}), 
while the red and blue dashed lines are the contributions from 
$\delta T_{\mathrm{c}}$ and $\delta l$ pinnings, respectively.
}
\label{fig:6}
\end{center}
\end{figure}

\subsection{Weak collective pinning}

It has been established that 
$J_{\mathrm{c}}$ in bulk FTS can be described by the weak collective pinning theory~\cite{Blatter1994}.
As shown in Fig.~\ref{fig:5}(b), 
$J_{\mathrm{c}}$ at 1~T in bulk FTS is very close to that in our thin-film devices with 
$d \gtrsim 100$~nm. Therefore, we
analyze $J_{\mathrm{c}}\left( T \right)$ using
the weak collective pinning theory, as discussed 
for bulk FTS in previous works~\cite{Das2011, Bonura2012, Galluzzi2020, Sun2013, Tamegai2016}. 
In the theoretical approach proposed by Griessen \textit{et al}.~\cite{Griessen1994}, 
$J_{\mathrm{c}}\left( T \right)$ in the moderately low field region, where 
the strong pinning can be ignored and the motion of single vortex is essential,
is given as
\begin{align}
J_{\mathrm{c}}^{\delta l}\left( t \right) =J_{\mathrm{c0}}^{\delta l}\left( 1-t^2 \right)^{5/2}\left( 1+t^2 \right)^{-1/2} \label{eq:2}
\end{align}
for $\delta l$ pinning and 
\begin{align}
J_{\mathrm{c}}^{\delta T_{\mathrm{c}}}\left( t \right)=J_{\mathrm{c0}}^{\delta T_{\mathrm{c}}}\left( 1-t^2 \right)^{7/6}\left( 1+t^2 \right)^{5/6} \label{eq:3}
\end{align}
for $\delta T_{\mathrm{c}}$ pinning; 
$ t $ is the reduced temperature ($ t=T/T_{\mathrm{c}} $). 
$ J_{\mathrm{c0}}^{\delta l}$ and $J_{\mathrm{c0}}^{\delta T_{\mathrm{c}}}$ 
are values at absolute zero temperature for $\delta l$ pinning and 
$\delta T_{\mathrm{c}}$ pinning, respectively. 
If both pinning mechanisms coexist, 
$ J_{\mathrm{c}}$ can be written as:
\begin{align}
    J_{\mathrm{c}}\left( t \right)=J_{\mathrm{c}}^{\delta l}\left( t \right)+J_{\mathrm{c}}^{\delta T_{\mathrm{c}}}\left( t \right).  \label{eq:4}
\end{align}

In Fig.~\ref{fig:3}(b), we show the best fit with Eqs.~(\ref{eq:2})--(\ref{eq:4}) for 
the 95~nm thick FTS device. 
We use two fitting parameters 
($J_{\mathrm{c0}}^{\delta l}$ and $J_{\mathrm{c0}}^{\delta T_{\mathrm{c}}}$) 
and obtain $J_{\mathrm{c0}}^{\delta l}=0\pm 0.06$~MA/cm$^2$, 
$J_{\mathrm{c0}}^{\delta T_{\mathrm{c}}}=0.77\pm 0.03$~MA/cm$^{2}$, respectively.
It is obvious that $J_{\mathrm{c0}}^{\delta T_{\mathrm{c}}}$ is much larger 
than $J_{\mathrm{c0}}^{\delta l} $, indicating that  
the $\delta T_{\mathrm{c}}$ pinning is much more dominant. 
In Fig.~\ref{fig:6}(a), we show the thickness dependence of 
$\eta^{\delta l}\,(\equiv J_{\mathrm{c0}}^{\delta l}/J_{\mathrm{c0}})$ 
and $\eta^{\delta T_{\mathrm{c}}}\,(\equiv J_{\mathrm{c0}}^{\delta T_{\mathrm{c}}}/J_{\mathrm{c0}})$
obtained from fitting,
where $J_{\mathrm{c0}}=J_{\mathrm{c0}}^{\delta l}+J_{\mathrm{c0}}^{\delta T_{\mathrm{c}}}$ 
is the total critical current density at $T=0$. 
As mentioned in the previous subsection, 
the fitting with Eqs.~(\ref{eq:2})--(\ref{eq:4}) works even for the strong pinning region. 
In the present $d$ range, 
$\eta^{\delta T_{\mathrm{c}}}$ is much larger than $ \eta^{\delta l} $, 
indicating that the $\delta T_{\mathrm{c}}$ pinning is more dominant. 
With increasing $d$, the proportion of $ \eta^{\delta l} $ becomes larger and exceeds that 
of $\eta^{\delta T_{\mathrm{c}}}$~\cite{Sun2013} in bulk FTS. 
This fact suggests that a crossover from the strong pinning to the weak collective pinning 
occurs at $d \approx 100$~nm.

Let us discuss the reason why the $\delta T_{\mathrm{c}}$ 
pinning is dominant in our thin-film devices. 
One possibility is the existence of excess Fe atoms.
In most previous works on bulk FTS~\cite{Das2011, Bonura2012, Galluzzi2020} 
where as-grown samples were used,
the $\delta l$ pinning is dominant.  
In such a case, a small amount of Fe impurities would be in between 
the two chalcogen layers.
To remove the Fe impurities, 
Sun~\textit{et al.} performed oxygen annealing for FTS and found that 
both the $\delta l$ and the $\delta T_{\mathrm{c}}$ pinnings coexist~\cite{Sun2013}. 
As mentioned in Sec.~II, 
excess Fe (i.e., atomic-scale) impurities would be segregated at the surface of FTS 
by annealing it under an oxygen flow. 
This results in the reduction of the $\delta l$ pinning contribution. 
To confirm the above scenario, we fabricated as-grown FTS thin-film devices and 
measured the temperature dependence of the critical current density 
in the same way as for annealed FTS devices.
Figure~\ref{fig:6}(b) shows the temperature dependence of 
$ J_{\mathrm{c}}\left( T \right) $ for a 105~nm thick as-grown FTS device
and also the result of fitting with Eqs.~(\ref{eq:2})--(\ref{eq:4}). 
When we measured $I$-$V$ curves for as-grown devices, the critical current 
changed with every measurement. 
Therefore, the data points in Fig.~\ref{fig:6}(b) are scattered compared with 
those for the annealed FTS device shown in Fig.~\ref{fig:3}(b). 
From the fitting with Eqs.~(\ref{eq:2})--(\ref{eq:4}), 
$ \eta^{\delta T_{\mathrm{c}}} $ 
and $ \eta^{\delta l} $ can be evaluated, as indicated by the triangle points in Fig.~\ref{fig:6}(a). 
The contribution of $\delta l$ pinning is comparable to 
that of $\delta T_{\mathrm{c}}$ pinning, even at $d \approx 100$~nm where 
the latter is much more dominant than the former in the annealed devices. 
This clearly shows that the excess Fe plays an important role 
in the pinning mechanism in FTS.
We note that even in as-grown FTS thin-film devices, 
the contribution of $ \delta T_{\mathrm{c}} $ pinning is larger than that of the annealed bulk FTS sample.

Another possibility is the similarity between $\delta T_{\mathrm{c}}$ pinning and strong pinning. 
In the case of strong pinning at nanometer-sized defects, 
the $\delta l$ pinning is less important than the $\delta T_{\mathrm{c}}$ pinning, 
as pointed out in Ref.~\cite{Beek2002}. 
This is because the scattering cross section of quasiparticles at atomic-scale defects is 
negligibly small compared with that at large-sized defects. 
In thin-film FTS devices~\cite{Zhu2013}, the inhomogeneous distribution of Te would be more essential.
Therefore, the contribution of the $\delta T_{\mathrm{c}}$ pinning is much more dominant 
in annealed FTS thin-film devices and is still large even in as-grown FTS thin-film devices 
which contain many atomic-scale defects.
In bulk FTS, on the other hand, 
quasiparticles may scatter at atomic-scale defects and thus the $\delta l$ pinning 
can be dominant because the path of the superconducting state is robust
and the inhomogeneous distribution is not crucial~\cite{Yue2016}.

\section{Conclusion}

We have studied the critical current density $J_{\mathrm{c}}$ in FeTe$ _{0.6} $Se$ _{0.4}$ (FTS) devices 
with thicknesses $d$ from 30 to 150~nm by means of electric transport measurements. 
Below $d \approx 70$~nm, $J_{\mathrm{c}}$ 
does not depend on the applied magnetic field and linearly increases with increasing $d$. 
This result is consistent with the strong pinning theory. 
Above $d \approx 100$~nm, on the other hand, $J_{\mathrm{c}}$ decreases with increasing $d$ 
and becomes comparable to that for bulk FTS, suggesting  
that the thicker film region corresponds to the weak collective pinning region. 
From these experimental results, 
a crossover from the strong pinning to the weak collective pinning 
is realized at $d^{*} = 70 \sim 100$~nm at a fixed magnetic field of 1~T. 
This value can be explained by the theoretical model where 
a defect size is comparable to the inhomogeneity of Te ($\approx 10$~nm) in FTS. 
Moreover, the $\delta T_{\mathrm{c}}$ pinning is much more dominant than 
the $\delta l$ pinning in our thin-film devices. 
This is partly due to the fact that our FTS have been annealed 
under an oxygen flow before the exfoliation, 
resulting in a reduction of excess Fe (i.e., atomic-size impurities) 
in between the two chalcogen layers.

\begin{acknowledgments}
This work was supported by
JSPS KAKENHI (Grants No. JP16H05964, No. JP17K18756, No. JP19K21850,
No. JP19H00656, No. JP19H05826, No. JP20H02557, No. JP20J20229, and No. JP21J20477), the Mazda Foundation, 
the Shimadzu Science
Foundation, the Yazaki Memorial Foundation for Science and Technology, the SCAT
Foundation, the Murata Science Foundation, Toyota Riken Scholar, 
the Kato Foundationfor Promotion of Science, and the Asahi Glass foundation.
\end{acknowledgments}



\end{document}